# Treatment plan comparison of Linac step and shoot, Tomotherapy, RapidArc, and Proton therapy for prostate cancer using dosimetrical and biological index

Suk Lee, Yuan Jie Cao, Kyung Hwan Chang, Jang Bo Shim, Kwang Hyeon Kim, Nam Kwon Lee, Young Je Park, Chul Yong Kim

*Department of Radiation Oncology, College of Medicine, Korea University, Seoul, 136-705*

Sam Ju Cho

*Department of Radiation Oncology, College of Medicine, Yonsei Cancer Center, Yonsei University, Seoul, 135-720*

Sang Hoon Lee

*Department of Radiation Oncology, Cheil General Hospital & Women's Healthcare Center, Kwandong University College of Medicine, Seoul, 100-380*

Chul Kee Min, Woo Chul Kim, Kwang Hwan Cho

*Department of Radiation Oncology, Soonchunhyang University Hospital, Bucheon 420-767*

Hyun Do Huh

*Department of Radiation Oncology, Inha University Hospital, Incheon, 400-701*

Sangwook Lim

*Department of Radiation Oncology, College of Medicine, Kosin University, Busan 602-702*

Dongho Shin

*Proton Therapy Center, National Cancer Center, Goyang, 410-769*

The purpose of this study was to use various dosimetrical indices to determine the best IMRT modality technique for treating patients with prostate cancer. Ten patients with prostate cancer were included in this study. Intensity modulated radiation therapy plans were designed to include different modalities, including the linac step and shoot, Tomotherapy, RapidArc, and Proton systems. Various dosimetrical indices, like the prescription isodose to target volume (PITV) ratio, conformity index (CI), homogeneity index (HI), target coverage index (TCI), modified dose homogeneity index (MHI), conformation number (CN), critical organ scoring index (COSI), and quality factor (QF) were determined to compare the different treatment plans. Biological indices such as the generalized equivalent uniform dose (gEUD), based tumor control probability (TCP), and normal tissue complication probability (NTCP) were also calculated and used to compare the treatment plans. The RapidArc plan attained better PTV coverage, as evidenced by its superior PITV, CI, TCI, MHI, and CN values. Regarding OARs, proton therapy exhibited superior dose sparing for the rectum and bowel in low dose volumes, whereas the Tomotherapy and RapidArc plans achieved better dose sparing in high dose volumes. The QF scores showed no significant difference among these plans (p=0.701). The average TCPs for prostate tumors in the RapidArc, Linac, and Proton plans were higher than the average TCP for Tomotherapy (98.79%, 98.76%, and 98.75% vs. 98.70%, respectively). Regarding the rectum NTCP, RapidArc showed the most favorable result (0.09%), whereas Linac resulted in the best bladder NTCP (0.08%).

E-mail: sukmp@korea.ac.kr

Fax: + 82-2-927-1419

# I. INTRODUCTION

Prostate cancer is the most common cancer worldwide; moreover, the incidence of prostate cancer has been rapidly increasing in all industrialized nations [1]. Since the introduction of advanced radiation therapy (RT) techniques such as intensity-modulated radiation therapy (IMRT), several studies have reported that IMRT yields excellent results and is particularly suited to the treatment of prostate cancer. This is because IMRT not only enables dose escalation in the prostate, but also simultaneously spares normal tissues such as the bladder and rectum [2-4]. Moreover, IMRT results in significantly less toxicity and significantly more prostate dose escalation compared with 3 dimensional conformal radiotherapy (3DCRT) plans [5-7]. Therefore, IMRT is being increasingly used in the treatment of prostate cancer.

Various modalities of IMRT delivery systems have been developed, such as linear accelerator based-IMRT systems, the Tomotherapy® system (Hi-Art system, Accuray Inc., Madison, WI, USA), the RapidArc system (Varian Medical System Inc., Palo Alto, CA, USA), the CyberKnife system (Accuray Inc., Sunnyvale, CA, USA), and proton systems.

Generally, IMRT delivery systems are classified into two different categories: fixed-gantry and arc-based. Fixed-gantry IMRT systems use a number of fixed beam directions with a multi-leaf collimator; these systems are based on the linear accelerator (LINAC). Step-and-shoot and dynamic (sliding window) IMRT techniques are some of these systems. One notable example of arc-based IMRT is the Tomotherapy® system, which can be divided into the TomoDirect and TomoHelical IMRT techniques [8-10]. In the step-and-shoot technique, a radiation beam is produced by the superposition of a number of static segments. The leaves of the MLC stay stationary during beam delivery; moreover, the beam is off when the MLC leaves and travels from one segment to another [11, 12]. The TomoDirect system allows the beam irradiation to be focused at discrete angles with a fixed gantry, which is particularly suited to certain clinical situations in which the beam arrangement is constrained to a limited number of pre-established directions using a continuous couch and MLC movement[13-15]. The TomoHelical system is a unique beam delivery technique in which beam irradiation can be generated at any gantry angle, coupled with continuous rotation of the gantry around the patient [16]. The CyberKnife is a novel beam delivery system that is well suited for treating the prostate. This system consists of a 6 MV linear accelerator that is mounted on a robotic arm, a pair of X-ray imaging systems, and several hundred noncoplanar beams [17, 18]. Proton therapy is also increasingly being used to treat prostate cancer. Moreover, proton therapy appears to result in significant dose savings to critical organs; proton therapy also results in decreased treatment toxicity and improved treatment efficacy due to the sharp characteristic Bragg peak of proton beams [19-21].

Several studies have compared treatment plans using different IMRT modalities by determining their dosimetric indices for prostate cases [17, 19, 22]. Ceylan et al. demonstrated that the CyberKnife can deliver high doses to the prostate compared with LINAC-based IMRT plans; however, the PTV coverage between these two modalities is not very significant [17]. Schwarz et al. compared helical tomotherapy and intensity modulated proton therapy (IMPT) techniques for treating prostate cancer. This study confirmed that IMPT provides better savings to OARs, in particular the bladder and penile bulb, compared with helical tomotherapy. However, both of these modalities offer good target coverage [22]. Similarly, Cotter et al. compared proton radiotherapy versus IMRT for pediatric prostate rhabdomyosarcoma. This study provides evidence of significant dose savings to normal structures with proton radiotherapy compared to IMRT [19]. As described above, several studies have compared the results of various IMRT delivery techniques. However, studies comparing prostate cancer outcomes for treatment plans using the Linac step and shoot technique, Tomotherapy, RapidArc, or proton modalities are limited.

The purpose of this study was thus to determine the best modality out of four possible modalities for treating patients with prostate cancer. To this end, we compared the dosimetric physical indices, EUD-based TCP scores, and EUD-based NTCP scores for different IMRT plans used to treat ten different patients with prostate cancer.

## II. Materials and Methods

### A. Patient characteristics

Ten consecutive patients treated with tomotherapy for prostate cancer at our institution were included in this study. All patients were male, with 8 patients aged older than 65 years. All patients wre diagnosed as clinical stage II or III prostate cancer and no patients received nodal irradiation.

### B. Planning CT scans and contouring PTVs and OARs

All patients underwent computed tomography (CT) scans for treatment planning on a CT simulator (Philips Medical System, Eindhoven, The Netherlands). Computed tomography images were acquired using a CT Simulator with 2 mm axial slice thickness. All CT data were transferred to a treatment planning system (TPS, Eclipse version 8.9, Varian Medical System Inc., Palo Alto, CA) with inverse planning. All clinical target volumes (CTVs), planning target volumes (PTVs), and organs at risk (OARs) were contoured on TPS. Organs considered to be at risk included the bladder, rectum, hip joint, and bone marrow. The CTV was defined as the region of prostate cancer. The PTV was determined by adding a 2 mm margin to the CTV in order to compensate for potential movements such as those occurring during treatment setup and/or breathing motions.

### C. Dose description

The prescription dose for each patient was designed so that 95% of the PTV received at least 77 Gy, with 35 fractions of 1.8 Gy administered daily. Dose constraints for OARs were determined based on the method described by Emami [23] and QUANTEC data [24]. The following dose constraints for OARs were used: mean dose for the bladder, 40 Gy; mean dose for the rectum, 40 Gy; mean dose for the hip joint, 50 Gy; mean dose for the bone marrow, 50 Gy.

### D. Planning techniques

#### D.1. LINAC-based IMRT treatment plan

A linear accelerator (Clinac iX, Varian Medical System Inc., Palo Alto, CA, USA) equipped with a 6 MV photon beam was used for step-and-shoot IMRT planning. Step-and-shoot IMRT treatment plans were generated in Eclipse TPS (Version 8.0, Varian Medical System Inc., Palo Alto, CA, USA). Beam modulation was performed using a 120-leaf MLC. The pencil beam algorithm was used to calculate the dose for each patient.

**D.2. TomoHelical IMRT plan**

Planning CT data with RT structures were transferred from Eclipse TPS to Pinnacle TPS; the CT data were then imported into the TomoTherapy Planning station (Hi-Art version 1.1.1, Accuray Inc., Madison, WI, USA) to generate the TomoHelical IMRT plans. The field width, pitch, and modulation factor values were 1.05 cm, 0.287, and 2.5, respectively. A calculation grid was applied according to standard procedures. The pencil beam algorithm was used for dose calculation.

**D.3. RapidArc plan**

To generate RapidArc plans, the Eclipse treatment planning system (Version 8.9, Varian Medical System Inc., Palo Alto, CA, USA) was used. The AAA algorithm was used for dose calculation

**D.4. Proton plan**

For proton treatment planning, CT data were imported into the proton treatment planning system (Eclipse Version 8.03, Varian Medial System Inc., Palo Alto, CA, USA). All proton treatment plans used one beam and the passive scattering mode. The gantry angle was 270°, and the AAA algorithm was used for dose calculation.

**E. Treatment plan analysis**

**E.1. Dosimetrical index**

Several quantitative evaluation tools were used to compare plans with one another. The following aspects of the plans were compared: prescription isodose to target volume (PITV) ratio, homogeneity index (HI), conformity index (CI), target coverage index (TCI), modified dose homogeneity index (MHI), conformity number (CN) for the PTV, maximum dose, mean dose, dose volume histogram (DVH), and critical organ scoring index (COSI) for the OAR. The PITV ratio was obtained by dividing the prescription isodose surface volume by the target volume [25]. The CI, which is defined as the ratio of the target volume and the volume inside the isodose surface that corresponds with the prescription dose, is generally used to indicate the portion of the prescription dose that is delivered inside the PTV [26]. The HI is the ratio of the maximum dose delivered to the PTV to the prescription dose to the PTV. The TCI refers to the exact coverage of the PTV in a treatment plan for a given prescription dose [27]. The MHI is similar to the HI, except that it is expressed as the 95% dose coverage value divided by the 5% dose coverage value [27]. The CN is a relative measurement of the dosimetric target coverage and the sparing of normal tissues in a given treatment plan [28]. The CN is expressed as:

$$CN = TCI \times CI = \frac{PTV_{PD}}{PTV} \times \frac{PTV_{PD}}{PIV} \quad (1)$$

where $PTV_{PD}$ refers to the PTV coverage at the prescription dose and PIV represents the prescription isodose surface volume. The COSI index takes into account both the target coverage and the critical organ irradiation [29]. The main advantage of this index is its ability to distinguish between different critical organs. The COSI is expressed as:

$$COSI = 1 - \sum_{1}^{n} w_i \frac{V_i(OAR)_{>tol}}{TC} \quad (2)$$

where $V(OAR)_{>tol}$ is the fraction of the volume of the OAR that receives more than a predefined tolerance dose, and $TC_V$ is the volumetric target coverage, which is defined as the fractional volume of PTV covered by the prescribed isodose.

The modified COSI is expressed as:

$$mCOSI = \sum_{i=1}^{n} Wi \left[ \frac{COSI10 + COSI20 + \cdots + COSI80}{8} \right] \quad (3)$$

Although the COSI index focuses only on the OARs that receive high dose region volumes, the modified COSI considers both high dose regions and low dose regions.

The maximum dose, mean dose, and DVH were used to quantitatively evaluate the dose distributions in the rectum and bladder. The DVH index included $V_5$, $V_{10}$, $V_{20}$, $V_{30}$, $V_{40}$, $V_{50}$, $V_{60}$, and $V_{70}$.

**E.2. Quality factor (QF)**

A novel dosimetrical index that can evaluate the quality of the entire plan, named the quality factor

(QF), was introduced in this study. The QF of a plan can be analytically expressed as:

$$QF = [2.718 \exp(-\sum_{i=1}^{N} W_i X_i)] \quad (4)$$

In the above equation, Xi represents all of the PTV indices used in this study, including the PITV, CI, HI, TCI, MHI, CN, and COSI. The values of the weighting factor (Wi) can be adjusted between 0 and 1 for all relatively weighted indices for a user-defined number of indices (N). In this study, a weighting factor of 1 was used for all separate indices. Thus, the QF was mainly used to compare the conformity of plans throughout the various trials of a treatment [30].

**E.3. Biological index**

For radiobiological model-based plan evaluation, Niemierko's equivalent uniform dose (EUD)-based NTCP and TCP models were used [31, 32]. First, the DVHs from each plan were exported from the appropriate treatment planning system (TPS) for each modality. Then, the DVHs were imported into MATLAB version R2012a (The MathWorks, Inc., Natick, Massachusetts) for TCP and NTCP modeling analysis.

According to Neimierko's phenomenological model, the EUD is defined as:

$$EUD = [\sum_{i=1}(V_i EQD_i^a)]^{\frac{1}{a}} \quad (5)$$

where a is a unitless model parameter that is specific to the nominal structure of the tumor of interest, and $v_i$ is a unitless parameter that represents the i$^{th}$ partial volume receiving dose $D_i$ in Gy [31]. Since the relative volume of the whole structure of interest corresponds to 1, the sum of all the partial volumes $v_i$ will equal 1. Furthermore, in equation (4), the EQD is the biologically equivalent physical dose of 2 Gy and is defined as:

$$EQD = D \times \frac{\left(\frac{\alpha}{\beta} + \frac{D}{n_f}\right)}{\left(\frac{\alpha}{\beta} + 2\right)} \quad (6)$$

where $n_f$ and $d_f = D/n_f$ are the number of fractions and the dose per fraction size of the treatment course, respectively. In this equation, α/β is the tissue-specific linear quadratic (LQ) parameter of the organ being exposed.

Niemierko's TCP [31] is defined as:

$$\text{TCP} = \frac{1}{1+\left(\frac{TCD_{50}}{EUD}\right)^{\gamma_{50}}} \qquad (7)$$

where $TCD_{50}$ is the tumor dose to control 50% of the cancer cells when the tumor is homogeneously irradiated, and $\gamma_{50}$ is a unitless model parameter that is specific to the tumor of interest and describes the slope of the dose response curve.

Niemierko's NTCP [32] is defined as:

$$\text{NTCP} = \frac{1}{1+\left(\frac{TD_{50}}{EUD}\right)^{\gamma_{50}}} \qquad (8)$$

where $TD_{50}$ is the tolerance dose of a 50% complication rate at a specific time (e.g. 5 years in the Emami et al. normal tissue tolerance data [23]) for an entire organ of interest; this parameter also describes the slope of the dose response curve.

For the prostate tumors in this study, the EUD and TCP values were calculated using an α/β value of 1.2. For the OARs, the α/β values used were 3.9 and 8.0 for the rectum and bladder, respectively [31]. A complete list of the radiobiological parameters used in this study is shown in Table 5.

**D.3. Statistical analysis**

Data are reported as means ± SDs. To determine whether the differences between dosimetrical and biological indices were significant, the Kruskal-Wallis test and the Mann-Whitney test were performed. All calculations were performed using SPSS software, version 19.0. Differences were considered significant for *p* values < 0.05.

### III. Results

**A. Dosimetrical index**

**A.1. PTV**

All investigated indices for the PTVs are shown in Table 1, and are expressed as mean values ± standard deviations (SDs). Regarding the different PTV indices, the proton plan had relatively favorable results for the PITV and HI (0.999 and 1.056; $p = 0.008$ and $p = 0.002$, respectively) compared with other modalities. In contrast, the Linac and Rapidarc plans exhibited better results with respect to the CI and CN (0.953 and 0.905; $p = 0.028$ and $p = 0.038$, respectively). Furthermore, the Tomotherapy plan yielded a slightly better MHI value (0.960) compared with the others. However, the TCI and MHI indices were not significantly different among any of the modalities examined. The different indices are compared among these four modalities in table 3. The proton plan showed significant discrepancies in almost all PTV indices (P < 0.05). However, the TCI and MHI values for the proton plan were not significantly different from those of the tomotherapy, Linac, and Rapidarc plans (P > 0.05). The majority of the PTV indices were not significantly different among the tomotherapy, Linac, and Rapidarc plans, with the exception of the HI values for the Linac versus the Rapidarc plans (P = 0.006).

### A.2. OAR

All proton, Tomotherapy, Linac, and Rapidarc treatment plans met the criteria for normal tissue constraints in this study. The mean values of volume for the normal organs were 68.11 cm$^3$, 106.43 cm$^3$, 456.51 cm$^3$, 24.86 cm$^3$, and 578.83 cm$^3$ for the rectum, bladder, bowel, hip joint, and bone marrow, respectivel1. Detailed dosimetric indices for the OARs are shown in Table 2. Regarding the OAR indices, the proton plan showed relatively favorable results since it incurred lower dose volumes (49.35, 43.39, 39.39, 36.23, and 30.88 for RV5, RV10, RV15, RV20, and RV30; 72.01, 65.98, 61.82, 58.30, and 52.03 for BV5, BV10, BV15, BV20, and BV30, respectively). However, the differences between the proton plan and the other plans decreased when the indices for higher dose volumes (RV50, RV60, and RV70; BV50, BV60, and BV70) were examined, as shown in Table 4. For the maximum dose in both the rectum and the bladder, all modalities showed similar doses (78.20 Gy – 80.16 Gy for the rectum, $p = 0.116$; 78.75 Gy – 81.13 Gy for the bladder, $p = 0.002$). In contrast, arc therapies such as the Tomotherapy and Rapidarc approaches resulted in relatively larger percentages of volume in lower dose volumes (RV5 to RV 30); however, these differences were less apparent in higher dose volumes (RV50 to RV 70) (Table 4). The COSI index takes into account both the target coverage and the OAR dose (V10 to V80 in this study) in critical structures. As shown in Table 1, the Linac and Proton plan exhibited relatively favorable COSI results compared with the Rapidarc, and Tomotherapy plans (0.734 and 0.708 versus 0.698 and 0.687, respectively; $p$=0.630).

**B. Quality factor**

The quality factor takes into account all of the dosimetrical indices. As shown in Table 1 the Linac and Rapidarc plans had the most favorable overall results (1.066 ± 0.011 and 1.068 ± 0.080, respectively) compared with the proton and Tomotherapy plans (1.073 ± 0.015 and 1.080 ± 0.032, respectively) ($p$ =0.701).

**C. Biological index**

**C.1. PTV**

The average EUD values for the prostate tumors in the Linac and proton plans (85.029 ± 0.449 Gy and 84.766 ± 0.479 Gy, respectively) were slightly higher than those in the Tomotherapy and Rapidarc plans (83.339 ± 0.575 Gy and 84.686 ± 0.427 Gy, respectively), which were showed in table 5. Although these differences were slight, they were statistically significant ($p$<0.001). Furthermore, the average TCP values for the prostate tumors were significantly higher in the Linac and proton plans (98.903 ± 0.381% and 98.766 ± 0.027%, respectively) compared with the Tomotherapy and Rapidarc plans (98.680 ± 0.036% and 98.761 ± 0.025%, respectively) ($p$<0.001).

**C.2. Rectum**

The average EUD values for the rectum in the proton and Tomotherapy plans (54.997 ± 8.826 Gy and 55.590 ± 5.440 Gy, respectively) were slightly higher than the values for the Linac and Rapidarc plans (52.209 ± 4.702 Gy and 51.592 ± 3.555 Gy, respectively); however, these differences were not statistically significant ($p$=0.112). The average NTCP values for the rectum in the proton and Tomotherapy plans (1.094 ± 1.370% and 0.530 ± 0.422%, respectively) were higher than those in the Linac and Rapidarc plans (0.191 ± 0.167% and 0.129 ± 0.095%, respectively); however, these values were not significantly different ($p$=0.109) (table 5).

**C.3. Bladder**

The average EUD values for the bladder in the proton and Tomotherapy plans (45.337 ± 6.864 Gy and 41.044 ± 7.698 Gy, respectively) were slightly higher than those in the Linac and Rapidarc plans (34.348 ± 6.351 Gy and 36.886 ± 6.558 Gy, respectively) ($p$=0.016). The average NTCP values for the bladder in the proton and Tomotherapy plans (0.065 ± 0.144% and 0.017 ± 0.026%, respectively) showed in table 5, were higher than those in the Linac and Rapidarc plans (0.002 ± 0.005% and 0.004 ± 0.007%, respectively); these differences were statistically significant ($p$=0.049).

## IV. Discussion

Here we used both dosimetric indices and EUD-based TCP and NTCP models to compare various IMRT plans (conventional LINAC-based IMRT, TomoDirect, TomoHelical, RapidArc, and proton techniques) for treating prostate cancer. Both dosimetrical indices and biological models revealed that the Linac and Rapidarc plans were superior to the proton and Tomotherapy plans, at least in the context of prostate radiotherapy.

As shown in Table 1 and 3, the proton therapy plan provided a slightly greater PITV, and the dose conformity in this plan was almost equal to the PTV. However, the plans with other modalities also achieved the dosimetric criteria for the PTV. Rao et al. reported that the homogeneous dose distribution within the PTV was slightly increased compared with the VMAT [33]. In contrast, Pasquier et al. reported that the HT provided a more accurate HI value compared with the VAMT [34].

The inherent arc therapy nature of the Tomotherapy and Rapidarc plans enabled them to produce highly conformal dose distributions over the targets, as evidenced by our findings that these plans yielded better PTV indices in terms of the PITV, CI, TCI, MHI, and CN. In accordance with our findings, Poon et al. reported that the Rapidarc plan was more conformal than static field IMRT for PTVs in high risk prostate cases [35].

Conversely, due to the rotation that is inherent in arc therapy approaches, substantial volumes of normal tissues adjacent to the tumors received low doses of irradiation when the doses were spread over a span of 360°. As shown in Table 2, arc therapies like Tomotherapy and Rapidarc resulted in larger percentages of volume receiving radiation in lower target dose volumes (RV5 to RV 30). These differences were less apparent in higher dose volumes (RV50 to RV 70) (Table 4).

As shown in Table 1, the proton therapy plan yielded reduced doses to the bladder (RV5 to RV20) and the rectum (BV5 to BV30) compared with other plans, at least when the target volumes received low doses. Our results are similar to those reported in other treatment planning studies [19, 20, 36]. The Linac-based IMRT plan also significantly reduced the dose received by the rectum (RV30 to RV 70) and bladder (BV50 to BV70) compared with the other modalities. This result conflicts with a previous study by Rao et al., which reported that the IMRT and VMAT plans were not significantly different in their sparing of OARs [33]. This discrepancy might result from study-to-study treatment plan variations, or the ways in which the treatment plans were designed.

To identify the most balanced plan in terms of both PTV coverage and OAR sparing, we used the quality factor (QF) to evaluate the quality of each treatment plan [30]. This measurement takes into account all of the PTV and OAR indices used in this study, including the PITV, CI, HI, TCI, MHI, CN, and COSI. The values of the weighting factor can be adjusted between 0 and 1 for all relatively weighted indices for a user-defined number of indices. Thus, the QF is mainly used to compare the conformity of different plans in the various trials of a treatment. As shown in Table 1, the RapidArc and Linac-based plans exhibited slightly greater QFs compared with the other modalities; the Tomotherapy plan exhibited the lowest QF score.

The method that we used to evaluate the different treatment plans does have some inherent limitations. Since the PTV and OAR values were weighted equally, the importance of each factor in the total plan quality may not have been fully considered. To determine the most appropriate weighting value to use for determining the QF, additional studies with large sample sizes that also include clinical outcome information on certain treatment sites, in addition to more advanced statistical analysis, are needed.

For outcome-based radiobiological evaluation of treatment plans, we used Niemierko's EUD-based TCP and NTCP models [31, 32]. The largest difference in NTCP values for the rectum was between the proton and Rapidarc plans; Similarly, the NTCP values for the bladder were the most different between the proton and Linac plans. Currently, adequate clinical radiobiological modeling data on multimodality planning for prostate cancer are lacking. This may be due to the current standard practice of evaluating treatment plans using dosimetrical DVH parameters alone. Using the NTCP model, Widesott et al. compared the treatment planning results for treating prostate cancer with the Tomotherapy and proton IMRT techniques [37]. This study found that the proton plan yielded more favorable NTCP results for the rectum.

In this study, the treatment, dose calculation, and optimization time were not included in the QF and thus did not contribute to the plan quality metric. Rao et al. reported that one major advantage of the VMAT approach compared with the HT and IMRT plans is its fast beam delivery time [34]. Oliver et al. demonstrated that the RapidArc plan has the lowest beam delivery time compared with both the IMRT and Tomotherapy plans [38]. We confirmed that the proton plan can reduce the treatment time; however, the overall quality of the proton plan was not significantly different compared with the other modalities. This may be because the proton plan is not an IMRT plan technique, and it only uses one port for treatment. However, Rao et al. found that decreased treatment times can reduce the amount of intrafraction motion during radiation therapy, and can thus improve the overall biological effectiveness of a plan [26]. Therefore, future studies comparing plans with different modalities will incorporate treatment, dose calculation, and optimization time into the QF calculation.

## III. CONCLUSION

Here we used dosimetric indices, in addition to a plan evaluation scoring index developed in-house, to compare different IMRT treatment plans using conventional LINAC-based IMRT, the TomoHelical system, the RapidArc system, and proton techniques. We found that, regarding PTV coverage, arc therapies such as the RapidArc system can attain better target coverage and yielded better PITV, CI, TCI, MHI, and CN values. For OARs, proton therapy is a superior technique for achieving optimum dose sparing for the rectum and bowel in low dose volumes; however, arc therapies such as the Tomotherapy and RapidArc systems can achieve better dose sparing in high dose volumes. Similarly, Linac and Rapidarc showed advantages in terms of BED-based TCP and NTCP. Along these lines, considering both the dosimetrical and biological indices, the RapidArc and Linac IMRT approaches were superior to the other techniques for prostate cancer radiotherapy.

## ACKNOWLEDGEMENT

This work was supported by the Korea University Grant.

## REREFENCES

[1] Jemal A., Thomas A., Murray T., Thun M. CA Cancer J Clin. **52,** 23-47 (2002).

[2] Group I. M. R. T. C. W. Int J Radiat Oncol Biol Phys. **51,** 880-914 (2001).

[3] Luxton G., Hancock S. L., Boyer A. L. Int J Radiat Oncol Biol Phys. **59,** 267-84 (2004).

[4] Vlachaki M. T., Teslow T. N., Amosson C., Uy N. W., Ahmad S. Med Dosim. **30,** 69-75 (2005).

[5] De Meerleer G., Vakaet L., Meersschout S., et al. Int J Radiat Oncol Biol Phys. **60,** 777-87 (2004).

[6] Sethi A., Mohideen N., Leybovich L., Mulhall J. Int J Radiat Oncol Biol Phys. **55,** 970-8 (2003).

[7] Wang L., Hoban P., Paskalev K., et al. Med Dosim. **30,** 97-103 (2005).

[8] Chen Y., Chen Q., Chen M., Lu W. Med Phys. **38,** 3013-24 (2011).

[9] Verhey L. J. Semin Radiat Oncol. **9,** 78-98 (1999).

[10] Yu C. X. Phys Med Biol. **40,** 1435-49 (1995).

[11] Siochi R. A. Int J Radiat Oncol Biol Phys. **43,** 671-80 (1999).

[12] Xia P., Verhey L. J. Med Phys. **25,** 1424-34 (1998).

[13] Franco P., Catuzzo P., Cante D., et al. Tumori. **97,** 498-502 (2011).

[14] Kong C., Yu S., Cheung K., et al. Biomed Imaging Interv J. **8,** e14 (2012).

[15] Reynders T., Tournel K., De Coninck P., et al. Radiother Oncol. **93,** 71-9 (2009).

[16] Lee S S. J., Chang KH, Cao YJ. . Journal of the Korean Physical Society. **60,** 1961-6 (2012).

[17] Ceylan C., Kucuk N., Bas Ayata H., Guden M., Engin K. Rep Pract Oncol Radiother. **15,** 181-9 (2010).

[18] Katz A. J. Technol Cancer Res Treat. **9,** 463-72 (2010).

[19] Cotter S. E., Herrup D. A., Friedmann A., et al. Int J Radiat Oncol Biol Phys. **81,** 1367-73 (2011).

[20] Fontenot J. D., Lee A. K., Newhauser W. D. Int J Radiat Oncol Biol Phys. **74,** 616-22 (2009).

[21] Vargas C., Fryer A., Mahajan C., et al. Int J Radiat Oncol Biol Phys. **70,** 744-51 (2008).

[22] Schwarz M., Pierelli A., Fiorino C., et al. Radiother Oncol. **98,** 74-80 (2011).

[23] Emami B., Lyman J., Brown A., et al. Int J Radiat Oncol Biol Phys. **21,** 109-22 (1991).

[24] Bentzen S. M., Constine L. S., Deasy J. O., et al. Int J Radiat Oncol Biol Phys. **76,** S3-9 (2010).

[25] Shaw E., Kline R., Gillin M., et al. Int J Radiat Oncol Biol Phys. **27,** 1231-9 (1993).

[26] Knoos T., Kristensen I., Nilsson P. Int J Radiat Oncol Biol Phys. **42,** 1169-76 (1998).

[27] Yoon M., Park S. Y., Shin D., et al. J Appl Clin Med Phys. **8,** 9-17 (2007).

[28] van't Riet A., Mak A. C., Moerland M. A., Elders L. H., van der Zee W. Int J Radiat Oncol Biol Phys. **37,** 731-6 (1997).

[29] Menhel J., Levin D., Alezra D., Symon Z., Pfeffer R. Phys Med Biol. **51,** 5363-75 (2006).

[30] Pyakuryal A., Myint W. K., Gopalakrishnan M., Jang S., Logemann J. A., Mittal B. B. J Appl Clin

Med Phys. **11,** 3013 (2010).

[31] Gay H. A., Niemierko A. Phys Med. **23,** 115-25 (2007).

[32] Luxton G., Keall P. J., King C. R. Phys Med Biol. **53,** 23-36 (2008).

[33] Pasquier D., Cavillon F., Lacornerie T., Touzeau C., Tresch E., Lartigau E. Int J Radiat Oncol Biol Phys. **85,** 549-54 (2013).

[34] Rao M., Yang W., Chen F., et al. Med Phys. **37,** 1350-9 (2010).

[35] Poon D. M., Kam M., Leung C. M., et al. Clin Oncol (R Coll Radiol). **25,** 706-12 (2013).

[36] Trofimov A., Nguyen P. L., Coen J. J., et al. Int J Radiat Oncol Biol Phys. **69,** 444-53 (2007).

[37] Widesott L., Pierelli A., Fiorino C., et al. Int J Radiat Oncol Biol Phys. **80,** 1589-600 (2011).

[38] Oliver M., Ansbacher W., Beckham W. A. J Appl Clin Med Phys. **10,** 3068 (2009).

Table 1. Dosimetrical indices of PTVs.

| Index | Proton Mean ± SD | | | Tomotherapy Mean ± SD | | | Linac Mean ± SD | | | Rapidarc Mean ± SD | | | p-value |
|---|---|---|---|---|---|---|---|---|---|---|---|---|---|
| PITV | 0.999 | ± | 0.001 | 0.997 | ± | 0.003 | 0.996 | ± | 0.002 | 0.997 | ± | 0.003 | 0.000 |
| CI | 0.951 | ± | 0.001 | 0.937 | ± | 0.034 | 0.953 | ± | 0.003 | 0.953 | ± | 0.003 | 0.008 |
| HI | 1.056 | ± | 0.008 | 1.073 | ± | 0.020 | 1.060 | ± | 0.012 | 1.081 | ± | 0.011 | 0.000 |
| TCI | 0.950 | ± | 0 | 0.934 | ± | 0.033 | 0.949 | ± | 0.002 | 0.950 | ± | 0 | 0.009 |
| MHI | 0.959 | ± | 0.005 | 0.960 | ± | 0.010 | 0.956 | ± | 0.004 | 0.957 | ± | 0.006 | 0.000 |
| CN | 0.903 | ± | 0.001 | 0.876 | ± | 0.061 | 0.905 | ± | 0.005 | 0.905 | ± | 0.003 | 0.007 |
| COSI | 0.708 | ± | 0.082 | 0.687 | ± | 0.095 | 0.734 | ± | 0.080 | 0.698 | ± | 0.080 | 0.630 |
| QF | 1.073 | ± | 0.015 | 1.080 | ± | 0.032 | 1.066 | ± | 0.011 | 1.068 | ± | 0.080 | 0.701 |

Table 2. Dosimetrical indices of OARs.

| Index | Proton Mean ± SD | Tomotherapy Mean ± SD | Linac Mean ± SD | Rapidarc Mean ± SD | p-value |
|---|---|---|---|---|---|
| RV5 | 49.515 ± 18.543 | 77.775 ± 17.254 | 70.513 ± 16.807 | 78.601 ± 17.342 | 0.007 |
| RV10 | 43.397 ± 17.680 | 65.075 ± 18.513 | 59.656 ± 18.793 | 64.872 ± 18.378 | 0.057 |
| RV15 | 39.396 ± 16.685 | 59.313 ± 18.704 | 51.441 ± 19.528 | 58.439 ± 17.937 | 0.115 |
| RV20 | 36.234 ± 15.881 | 51.147 ± 17.319 | 42.935 ± 17.873 | 48.361 ± 13.891 | 0.315 |
| RV30 | 30.887 ± 14.309 | 35.845 ± 13.506 | 29.191 ± 13.183 | 32.527 ± 8.476 | 0.701 |
| RV50 | 21.387 ± 11.171 | 15.786 ± 6.947 | 14.045 ± 6.605 | 14.057 ± 4.558 | 0.402 |
| RV60 | 16.101 ± 9.159 | 9.785 ± 5.464 | 9.112 ± 4.541 | 8.372 ± 3.243 | 0.243 |
| RV70 | 9.266 ± 6.365 | 4.219 ± 3.131 | 4.059 ± 2.312 | 3.305 ± 1.565 | 0.148 |
| RMax Dose | 78.217 ± 2.246 | 78.442 ± 1.618 | 78.201 ± 1.830 | 78.861 ± 2.205 | 0.824 |
| BV5 | 72.010 ± 13.268 | 95.150 ± 7.413 | 89.672 ± 11.313 | 97.282 ± 5.017 | 0.000 |
| BV10 | 65.988 ± 13.708 | 83.500 ± 15.289 | 76.496 ± 16.346 | 84.304 ± 14.216 | 0.082 |
| BV15 | 61.829 ± 13.668 | 77.685 ± 19.143 | 66.046 ± 17.614 | 77.773 ± 18.676 | 0.203 |
| BV20 | 58.303 ± 13.596 | 72.374 ± 21.738 | 56.807 ± 16.710 | 71.374 ± 20.724 | 0.192 |
| BV30 | 52.037 ± 13.268 | 56.565 ± 21.449 | 42.311 ± 13.750 | 54.565 ± 19.329 | 0.309 |

| | | | | | | | | | | | | |
|---|---|---|---|---|---|---|---|---|---|---|---|---|
| BV50 | 38.812 | ± | 10.783 | 23.341 | ± | 9.871 | 21.131 | ± | 8.267 | 22.584 | ± | 8.608 | 0.005 |
| BV60 | 31.155 | ± | 9.193 | 13.121 | ± | 6.632 | 12.712 | ± | 5.354 | 12.511 | ± | 5.514 | <0.001 |
| BV70 | 20.864 | ± | 7.074 | 5.372 | ± | 3.924 | 4.960 | ± | 2.824 | 4.939 | ± | 2.620 | <0.001 |
| BMax Dose | 79.811 | ± | 0.687 | 79.101 | ± | 1.218 | 78.756 | ± | 0.877 | 79.555 | ± | 1.417 | 0.089 |

Abbreviations: RVX = relative rectal volume of receiving higher than X Gy, BVX = relative bladder volume of receiving higher than X Gy, Rmax dose = maximum dose to rectum, BMax dose = maximum dose to bladder.

Table 3. Comparison of PTV indices for each modality.

| Index | Proton p-value | | | | Tomotherapy p-value | | | | Linac p-value | | | |
|---|---|---|---|---|---|---|---|---|---|---|---|---|
| PITV | | | | | | | | | | | | |
| Tomotherapy | 0.019 | Proton | > | Tomotherapy | | | | | | | | |
| Linac | 0.001 | Proton | > | Linac | 0.472 | Tomotherapy | > | Linac | | | | |
| Rapidarc | 0.021 | Proton | > | Rapidarc | 0.970 | Tomotherapy | < | Rapidarc | 0.384 | Linac | < | Rapidarc |
| CI | | | | | | | | | | | | |
| Tomotherapy | 0.028 | Proton | > | Tomotherapy | | | | | | | | |
| Linac | 0.008 | Proton | < | Linac | 0.450 | Tomotherapy | < | Linac | | | | |
| Rapidarc | 0.025 | Proton | < | Rapidarc | 0.970 | Tomotherapy | < | Rapidarc | 0.496 | Linac | < | Rapidarc |
| HI | | | | | | | | | | | | |
| Tomotherapy | 0.028 | Proton | > | Tomotherapy | | | | | | | | |
| Linac | 0.325 | Proton | > | Linac | 0.096 | Tomotherapy | < | Linac | | | | |
| Rapidarc | 0.001 | Proton | > | Rapidarc | 0.448 | Tomotherapy | > | Rapidarc | 0.006 | Linac | > | Rapidarc |
| TCI | | | | | | | | | | | | |
| Tomotherapy | 0.136 | Proton | < | Tomotherapy | | | | | | | | |

| | | | | | | | | | | | |
|---|---|---|---|---|---|---|---|---|---|---|---|
| Linac | 0.317 | Proton | < | Linac | 0.328 | Tomotherapy | < | Linac | | | |
| Rapidarc | 0.543 | Proton | < | Rapidarc | 0.347 | Tomotherapy | < | Rapidarc | 0.957 | Linac | < | Rapidarc |

MHI

| | | | | | | | | | | | |
|---|---|---|---|---|---|---|---|---|---|---|---|
| Tomotherapy | 0.597 | Proton | > | Tomotherapy | | | | | | | |
| Linac | 0.290 | Proton | < | Linac | 0.571 | Tomotherapy | > | Linac | | | |
| Rapidarc | 0.226 | Proton | < | Rapidarc | 0.495 | Tomotherapy | > | Rapidarc | 0.910 | Linac | < | Rapidarc |

CN

| | | | | | | | | | | | |
|---|---|---|---|---|---|---|---|---|---|---|---|
| Tomotherapy | 0.031 | Proton | > | Tomotherapy | | | | | | | |
| Linac | 0.009 | Proton | < | Linac | 0.450 | Tomotherapy | < | Linac | | | |
| Rapidarc | 0.044 | Proton | < | Rapidarc | 0.970 | Tomotherapy | < | Rapidarc | 0.496 | Linac | < | Rapidarc |

COSI

| | | | | | | | | | | | |
|---|---|---|---|---|---|---|---|---|---|---|---|
| Tomotherapy | 0.003 | Proton | < | Tomotherapy | | | | | | | |
| Linac | 0.006 | Proton | < | Linac | 0.625 | Tomotherapy | > | Linac | | | |
| Rapidarc | 0.018 | Proton | < | Rapidarc | 0.347 | Tomotherapy | > | Rapidarc | 0.649 | Linac | < | Rapidarc |

QF

| | | | | | | | | | | | |
|---|---|---|---|---|---|---|---|---|---|---|---|
| Tomotherapy | 0.023 | Proton | > | Tomotherapy | | | | | | | |
| Linac | 0.009 | Proton | < | Linac | 0.059 | Tomotherapy | < | Linac | | | |
| Rapidarc | 0.000 | Proton | < | Rapidarc | 0.820 | Tomotherapy | < | Rapidarc | 0.001 | Linac | < | Rapidarc |

Table 4. Comparison of OAR indices for each modality.

| Index | Proton p-value | | | | Tomotherapy p-value | | | | Linac p-value | | | |
|---|---|---|---|---|---|---|---|---|---|---|---|---|
| RV5 | | | | | | | | | | | | |
| Tomotherapy | 0.005 | Proton | > | Tomotherapy | | | | | | | | |
| Linac | 0.028 | Proton | > | Linac | 0.226 | Tomotherapy | < | Linac | | | | |
| Rapidarc | 0.005 | Proton | > | Rapidarc | 0.879 | Tomotherapy | > | Rapidarc | 0.199 | Linac | > | Rapidarc |
| RV10 | | | | | | | | | | | | |
| Tomotherapy | 0.019 | Proton | > | Tomotherapy | | | | | | | | |
| Linac | 0.096 | Proton | > | Linac | 0.406 | Tomotherapy | < | Linac | | | | |
| Rapidarc | 0.023 | Proton | > | Rapidarc | 0.820 | Tomotherapy | < | Rapidarc | 0.450 | Linac | > | Rapidarc |
| RV15 | | | | | | | | | | | | |
| Tomotherapy | 0.041 | Proton | > | Tomotherapy | | | | | | | | |
| Linac | 0.326 | Proton | > | Linac | 0.257 | Tomotherapy | < | Linac | | | | |
| Rapidarc | 0.049 | Proton | > | Rapidarc | 0.879 | Tomotherapy | < | Rapidarc | 0.257 | Linac | > | Rapidarc |
| RV20 | | | | | | | | | | | | |
| Tomotherapy | 0.112 | Proton | > | Tomotherapy | | | | | | | | |

| | | | | | | | | | | | |
|---|---|---|---|---|---|---|---|---|---|---|---|
| Linac | 0.597 | Proton | > | Linac | 0.290 | Tomotherapy | < | Linac | | | |
| Rapidarc | 0.174 | Proton | > | Rapidarc | 0.820 | Tomotherapy | < | Rapidarc | 0.290 | Linac | > | Rapidarc |
| RV30 | | | | | | | | | | | |
| Tomotherapy | 0.597 | Proton | > | Tomotherapy | | | | | | | |
| Linac | 0.705 | Proton | < | Linac | 0.257 | Tomotherapy | < | Linac | | | |
| Rapidarc | 1.000 | Proton | > | Rapidarc | 0.762 | Tomotherapy | < | Rapidarc | 0.290 | Linac | > | Rapidarc |
| RV50 | | | | | | | | | | | |
| Tomotherapy | 0.257 | Proton | < | Tomotherapy | | | | | | | |
| Linac | 0.174 | Proton | < | Linac | 0.545 | Tomotherapy | < | Linac | | | |
| Rapidarc | 0.151 | Proton | < | Rapidarc | 0.705 | Tomotherapy | < | Rapidarc | 0.762 | Linac | > | Rapidarc |
| RV60 | | | | | | | | | | | |
| Tomotherapy | 0.151 | Proton | < | Tomotherapy | | | | | | | |
| Linac | 0.112 | Proton | < | Linac | 0.762 | Tomotherapy | < | Linac | | | |
| Rapidarc | 0.070 | Proton | < | Rapidarc | 0.705 | Tomotherapy | < | Rapidarc | 0.821 | Linac | < | Rapidarc |
| RV70 | | | | | | | | | | | |
| Tomotherapy | 0.096 | Proton | < | Tomotherapy | | | | | | | |
| Linac | 0.096 | Proton | < | Linac | 0.762 | Tomotherapy | < | Linac | | | |
| Rapidarc | 0.041 | Proton | < | Rapidarc | 0.820 | Tomotherapy | < | Rapidarc | 0.406 | Linac | < | Rapidarc |

RMax Dose

| | | | | | | | | | | |
|---|---|---|---|---|---|---|---|---|---|---|
| Tomotherapy | 0.406 | Proton | > | Tomotherapy | | | | | | |
| Linac | 0.910 | Proton | < | Linac | 0.449 | Tomotherapy | < | Linac | | |
| Rapidarc | 0.705 | Proton | > | Rapidarc | 0.596 | Tomotherapy | > | Rapidarc | 0.762 | Linac > Rapidarc |

BV5

| | | | | | | | | | | |
|---|---|---|---|---|---|---|---|---|---|---|
| Tomotherapy | 0.001 | Proton | > | Tomotherapy | | | | | | |
| Linac | 0.006 | Proton | > | Linac | 0.142 | Tomotherapy | < | Linac | | |
| Rapidarc | 0.000 | Proton | > | Rapidarc | 0.716 | Tomotherapy | > | Rapidarc | 0.045 | Linac > Rapidarc |

BV10

| | | | | | | | | | | |
|---|---|---|---|---|---|---|---|---|---|---|
| Tomotherapy | 0.041 | Proton | > | Tomotherapy | | | | | | |
| Linac | 0.151 | Proton | > | Linac | 0.289 | Tomotherapy | < | Linac | | |
| Rapidarc | 0.034 | Proton | > | Rapidarc | 0.848 | Tomotherapy | > | Rapidarc | 0.256 | Linac > Rapidarc |

BV15

| | | | | | | | | | | |
|---|---|---|---|---|---|---|---|---|---|---|
| Tomotherapy | 0.112 | Proton | > | Tomotherapy | | | | | | |
| Linac | 0.597 | Proton | > | Linac | 0.173 | Tomotherapy | < | Linac | | |
| Rapidarc | 0.112 | Proton | > | Rapidarc | 0.970 | Tomotherapy | > | Rapidarc | 0.174 | Linac > Rapidarc |

BV20

| | | | | |
|---|---|---|---|---|
| Tomotherapy | 0.131 | Proton | > | Tomotherapy |

| Linac | 0.705 | Proton | < | Linac | 0.131 | Tomotherapy | < | Linac | | | |
| Rapidarc | 0.131 | Proton | > | Rapidarc | 0.820 | Tomotherapy | < | Rapidarc | 0.131 | Linac | > | Rapidarc |

BV30

| Tomotherapy | 0.650 | Proton | > | Tomotherapy | | | | | | | |
| Linac | 0.082 | Proton | < | Linac | 0.151 | Tomotherapy | < | Linac | | | |
| Rapidarc | 0.705 | Proton | > | Rapidarc | 0.940 | Tomotherapy | < | Rapidarc | 0.174 | Linac | > | Rapidarc |

BV50

| Tomotherapy | 0.008 | Proton | < | Tomotherapy | | | | | | | |
| Linac | 0.003 | Proton | < | Linac | 0.821 | Tomotherapy | < | Linac | | | |
| Rapidarc | 0.002 | Proton | < | Rapidarc | 0.879 | Tomotherapy | < | Rapidarc | 0.880 | Linac | > | Rapidarc |

BV60

| Tomotherapy | 0.001 | Proton | < | Tomotherapy | | | | | | | |
| Linac | 0.000 | Proton | < | Linac | 0.940 | Tomotherapy | < | Linac | | | |
| Rapidarc | 0.001 | Proton | < | Rapidarc | 0.940 | Tomotherapy | < | Rapidarc | 0.940 | Linac | < | Rapidarc |

BV70

| Tomotherapy | 0.000 | Proton | < | Tomotherapy | | | | | | | |
| Linac | 0.000 | Proton | < | Linac | 0.940 | Tomotherapy | < | Linac | | | |
| Rapidarc | 0.000 | Proton | < | Rapidarc | 0.879 | Tomotherapy | < | Rapidarc | 0.940 | Linac | < | Rapidarc |

| | | | | | | | | | | | |
|---|---|---|---|---|---|---|---|---|---|---|---|
| BMax Dose | | | | | | | | | | | |
| Tomotherapy | 0.112 | Proton | < | Tomotherapy | | | | | | | |
| Linac | 0.026 | Proton | < | Linac | 0.326 | Tomotherapy | < | Linac | | | |
| Rapidarc | 0.821 | Proton | < | Rapidarc | 0.324 | Tomotherapy | > | Rapidarc | 0.070 | Linac | > | Rapidarc |

Abbreviations: RVX = relative rectal volume of receiving higher than X Gy, BVX = relative bladder volume of receiving higher than X Gy, Rmax dose = maximum dose to rectum, BMax dose = maximum dose to bladder.

Table 5. Radiobiological parameters used to calculate EUD-based TCP and NTCP scores.

| Organs | Indices | Proton Mean ± SD | Tomotherapy Mean ± SD | Linac Mean ± SD | Rapidarc Mean ± SD | p-value |
|---|---|---|---|---|---|---|
| PTV | EUD (Gy) | 84.766 ± 0.479 | 83.339 ± 0.575 | 85.029 ± 0.449 | 84.686 ± 0.427 | 0 |
|  | TCP (%) | 98.766 ± 0.027 | 98.68 ± 0.036 | 98.903 ± 0.381 | 98.761 ± 0.025 | 0 |
| Rectum | EUD (Gy) | 54.997 ± 8.826 | 55.59 ± 5.44 | 52.209 ± 4.702 | 51.592 ± 3.555 | 0.112 |
|  | NTCP (%) | 1.094 ± 1.37 | 0.53 ± 0.422 | 0.191 ± 0.167 | 0.129 ± 0.095 | 0.109 |
| Bladder | EUD (Gy) | 45.337 ± 6.864 | 41.044 ± 7.698 | 34.338 ± 6.351 | 36.886 ± 6.558 | 0.016 |
|  | NTCP (%) | 0.065 ± 0.144 | 0.017 ± 0.026 | 0.002 ± 0.005 | 0.004 ± 0.007 | 0.049 |

| Organ | Volume type | 100% Dpf | #f | a | $\gamma_{50}$ | $TD_{50}$ (Gy) | $TCD_{50}$ (Gy) | Dpf (Gy) | α/β (Gy) |
|---|---|---|---|---|---|---|---|---|---|
| PTV | Tumor | 1.9 | 35 | -10 | 1 | - | 28.34 | 2 | 1.2 |
| Rectum | Normal | 1.9 | 35 | 8.33 | 4 | 80 | - | 2 | 3.9 |
| Bladder | normal | 1.9 | 35 | 2 | 4 | 80 | - | 2 | 8 |

Abbreviations: 100% Dpf: 100% dose per fraction, #f: number of fractions, α/β: alpha-beta ratio, Dpf: dose per fraction, TD: tolerance dose, TCD: tumor control dose.